\documentclass[twocolumn]{aastex62}
\usepackage{graphicx}
\usepackage{amssymb}                    
\usepackage{ulem}
\usepackage{float}
\usepackage{rotating}
\usepackage{enumitem}
\usepackage{rotating}
\usepackage{makecell}
\usepackage{appendix}
\usepackage{fixmath}
\usepackage{epstopdf}
\usepackage{times} 
\usepackage{amsmath}
\usepackage{booktabs}

\definecolor{green1}{rgb}{0.2, 0.6, 0.1}
\definecolor{purp1}{rgb}{0.6, 0.1, 0.45}

\newcommand{\hic}{Hi-C\,2.1}
\newcommand{\ar}{AR\,12712}

\newcommand{\arc}{\ensuremath{^{\prime\prime}}}

\received{18 June 2021}
\accepted{27 June 2021}
\submitjournal{ApJ}

\shortauthors{Williams, Walsh, and Morgan (2021)}
\shorttitle{Loop Shape and Expansion}
\begin{document}

\title{The cross-sectional shape and height expansion of coronal loops: High-resolution Coronal Imager (Hi-C) analysis of AR\,12712}
\author{Thomas Williams}
\affil{Jeremiah Horrocks Institute, UCLan, Preston, PR1 2HE, UK}
\affil{Department of Physics, Abersytwyth University, Penglais, Aberystwyth, SY23 3BZ, UK}
\author{Robert W.\ Walsh}
\affil{Jeremiah Horrocks Institute, UCLan, Preston, PR1 2HE, UK}
\author{Huw Morgan}
\affil{Department of Physics, Abersytwyth University, Penglais, Aberystwyth, SY23 3BZ, UK}
\begin{abstract}
Coronal loop observations have existed for many decades yet the precise shape of these fundamental coronal structures is still widely debated since the discovery that they appear to undergo negligible expansion between their footpoints and apex. In this work a selection of eight EUV loops and their twenty-two sub-element strands are studied from the second successful flight of NASA's High resolution Coronal Imager (Hi-C\,2.1). Four of the loops correspond to open fan structures with the other four considered to be magnetically closed loops. Width analysis is performed on the loops and their sub-resolution strands using our method of fitting multiple Gaussian profiles to cross-sectional intensity slices. It is found that whilst the magnetically closed loops and their sub-element strands do not expand along their observable length, open fan structures may expand an additional 150\,\% of their initial width. Following recent work, the Pearson correlation coefficient between peak intensity and loop/strand width are found to be predominantly positively correlated for the loops ($\approx88$\,\%) and their sub-element strands ($\approx80$\,\%). These results align with the hypothesis of Klimchuk \& DeForest that loops and -- for the first time -- their sub-element strands have approximately circular cross-sectional profiles.
\end{abstract}

\section{Introduction}\label{sec:intro}
It has long been shown that coronal loops may have uniform cross sectional widths along their observable length \citep{klimchuk92,klimchuk00,watko00} where the cross-sectional profile is roughly circular. This behaviour is contrary to what one would expect; that is, flux tubes and thus coronal loops are expected to expand with height to maintain the pressure balance between the internal and external loop plasma due to gravitational stratification and decreasing magnetic field strength. This perceived circular cross-section may be explained by locally twisted flux tubes \citep{klimchuk2000,fuentes06} which have been demonstrated to undergo less expansion than untwisted structures \citep{mcclymont94}.

The findings of \citet{malanushenko13} offer an explanation for the perceived lack of expansion in loop observations that one would expect -- the loops may have non-circular cross sections such as an ellipse. As noted by \citet{klimchuk20}, if the expansion occurs preferentially along the line of sight, i.e. radially, rather than in the image plane, then no contradiction occurs. \citet{malanushenko13} propose that because the coronal plasma is optically thin, loops expanding along the line of sight will be brighter in the image plane, which causes a selection bias as the brightness will decrease less rapidly as a function of height due to the increasing line of sight depth towards the apex. Whilst it is possible that the cross-sectional aspect ratio may change along a loop's length, it is also possible that the aspect ratio remains fixed and merely rotates around a central axis, such as a twisting ribbon/rubber band \citep{mccarthy21}. However, if that were the case then a loop would experience localised brightening and dimming along its length. That is, if we consider the twisting of an oval cross section, then there is more (less) material along the line of sight when the semi-minor (major) cross-sectional axis is lined preferentially with the observer, which leads to brighter (dimmer) intensity in the 2D projection of observational data due to the increased (decreased) plasma along the line of sight.

To test this hypothesis, \citet{klimchuk20} analyse the relationship between peak intensity and cross-sectional width of twenty coronal loops within the 193\,\AA\ High Resolution Coronal Imager (Hi-C; \citealp{kobayashi14}) data-set whose lengths are $<60$\,Mm. They find that these two properties were either positively correlated or uncorrelated but not negatively correlated, suggesting the cross sections of the loops could be approximately circular. More recently, \citet{mccarthy21} utilised observations from two vantage points with 171\,\AA\ data from both Solar Dynamics Observatory's (SDO) Atmospheric Imaging Assembly (AIA; \citealp{lemen12}) and Solar Terrestrial Relation Observatory's (STEREO) Extreme Ultraviolet Imager (EUVI; \citealp{howard08}). From their sample of 151 loops, they identified thirteen as being suitable for width analysis. Utilising diameter-diameter measurements and peak intensity \textit{vs}. loop width graphs, \citet{mccarthy21} deduce that four of the thirteen loops may be elliptical with the remaining nine loops being approximately circular in shape. However, the diameter-diameter measurements are unlikely to be reliable given the lower resolution of EUVI, especially when AIA has been shown to be incapable of resolving -- or in some instances detecting -- narrower loops \citep{williams20a}.

As is discussed in \citet{klimchuk00}, the lack of cross-sectional expansion in loop observations have important implications for coronal heating models. The first is that energy may be deposited in an axial symmetric manner at the same spatial scales as the diameter of monolithic coronal loops/strands (full-width at half-maximum (FWHM) $\lesssim200$\,km; \citealp{williams20a,williams20b}). Alternatively, the energy deposited within strands and loops occur at spatial scales much smaller than the diameters of these structures -- perhaps on spatial scales as small as 15\,km \citep{peter13} -- and is then transported orthogonal to the loop abscissa in an axisymmetric manner. One popular mechanism for the latter scenario is the nanoflare heating model \citep{parker83,parker88} whereby field lines become braided and undergo small-scale magnetic reconnection due to photospheric motion at the footpoints.

In light of recent Hi-C findings indicating that monolithic loops may have approximately circular cross sections that exhibit no expansion along their observable length \citep{klimchuk20}, this paper employs the methods used in previous studies (\S\,2) to examine the spatial scales along the loop envelope \citep{williams20a} and individual loop sub-element strands \citep{williams20b} of eight distinct coronal structures. In \S\,3, the relationship between loop width and their peak intensities are explored to classify whether their cross sections may be approximately circular \citep{klimchuk20} or oval such as a twisted ribbon \citep{malanushenko13,mccarthy21}. For the first time, this study also extends this analysis to the sub-element strands contained within coronal loops and a discussion of these results in the context of current literature and the authors' concluding remarks are presented in \S\,4.

\vspace{1cm}
\section{Data Preparation and Analysis Method}\label{sec:data}
On 29\textsuperscript{th} May 2018 at 18:54 UT, \hic\ was successfully relaunched from the White Sands Missile Range, NM, USA, capturing high-resolution data (2k$\times$2k pixels; $4.4^\prime\times4.4^\prime$ field of view) of target active region \ar\ in EUV emission of wavelength 172\,\AA\ (dominated by Fe {\sc ix} emission $\approx0.8$\,MK) with a plate scale of 0.129\arc. During the flight \hic\ captured 78 images with a 2s exposure time and a 4.4s cadence between 18:56 and 19:02 UT. Full details on the \hic\ instrument can be found in \citet{instpaper}.

\begin{figure*}
\centerline{\includegraphics[width=\textwidth]{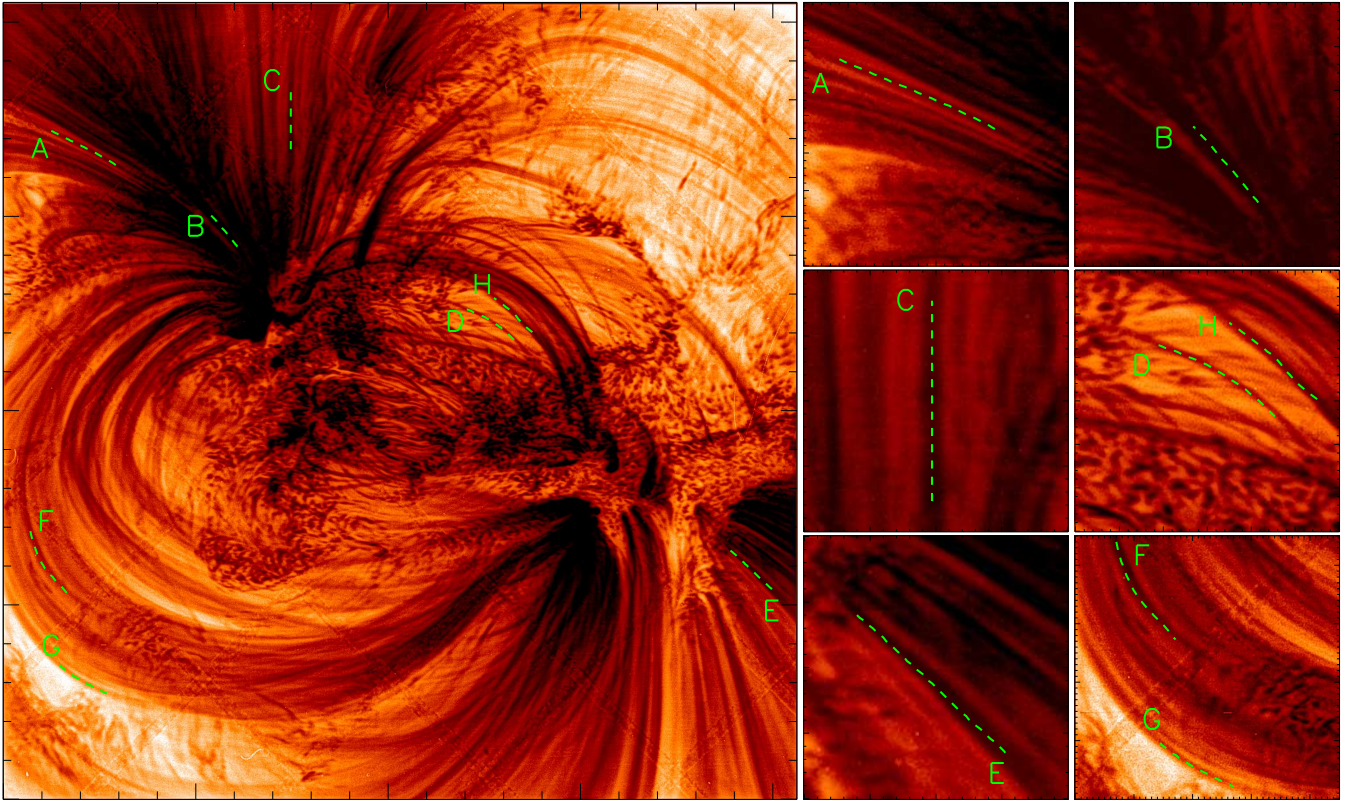}}
\caption{Reverse colour image showing the \hic\ field of view (\textit{left}) which has been time-averaged for $\approx60$~s and then, for the purpose of this figure only, sharpened with Multi-Scale Gaussian Normalisation \citep[MGN]{morgandruckmuller14}. The segments of the five loops analysed (A - H) are then shown in more detail in the panels on the \textit{right}.}
\label{fig:fov}
\end{figure*}
\subsection{Data-set Extraction and Background Subtraction}\label{sec:bkgd}
The basis of the sample data-set investigated here include a number of subsets from the ten higher-emission cross-section slices analysed by \citet{williams20a} plus nine other additional slices from within the \hic\ field of view (see Figure~\ref{fig:fov} where all data-set locations are indicated). In each case the resulting emission profile across the structures would indicate sub-structure strands that are not fully resolved.

Following the method outlined in \citep{williams20a}, the \hic\ data-set under consideration is time-averaged over a period $\approx60$\,s that is free from spacecraft jitter\footnote{A consequence of the instability experienced during the \hic\ flight is that ghosting of the mesh could not be avoided \citep{instpaper}. This leads to the diamond patterns across the entire \hic\ field-of-view, which are exaggerated when the data is enhanced with MGN (Figure\,\ref{fig:fov}).}. Each cross-section normal to each strand is taken to be 3-pixels deep and the background emission is then subtracted. As outlined in Figure~2 of \citet{williams20b}, this background subtraction is performed by firstly finding all the local minima of a slice, and interpolating through these values using a cubic spline \citep{yi15} to obtain a global trend (dashed blue line). The global trend is then subtracted from the intensity profile along the slice, leaving behind the background subtracted coronal strands (similar to \citealp{aschwanden11,williams20a}). Due to the large number of counts detected by \hic\, the Poisson error associated with these isolated coronal strands is minimal.

\subsection{Gaussian Fitting and FWHM Measurements}\label{sec:gaussfit}
The analysis method is based on the assumption that at rest, an isolated coronal strand element has an observed emission profile across its width and normal to the strand axis that is approximately Gaussian. It is important to note that as indicated by \citet{pontin17}, instantaneously coronal strands may not necessarily have a clear Gaussian cross-section. On the other hand, \citep{klimchuk20} have shown from Hi-C observations that coronal strands are likely to have Gaussian cross-sections. The data samples are time-averaged over $\approx60$\,s (the first 11 \hic\ frames) to average out variations over short timescales, which helps address this issue. Whilst no obvious signs of motion within the structures analysed are noticed in this 60\,s window, the authors acknowledge that as indicated by \citet{morton13}, small amplitude oscillations could be present which would lead to the measured widths being broader than the structural width due to the time integration performed.

The observed \hic\ intensity profile of a cross-sectional slice is reproduced by simultaneously fitting Gaussian profiles, the number of which is determined by the Akaike Information Criterion \citep[AIC]{akaike74} along with a corrective term (AICc) for small sample sizes. This is fully described in the Appendix of \citet{williams20b}. Subsequently, the full width at half maximum (FWHM) of the Gaussian profile is measured to provide an estimate of the possible width of the sub-structures likely present within the \hic\ data.

Thus, the method employed to fit Gaussian profiles to the observed \hic\ intensity is as follows. Firstly, the following expression for a Gaussian function, $Y_G$ is used:
\begin{equation}\label{eq:gaussian}
Y_{G} = A \exp\left(\frac{-(x-x_{p})^2}{2W^2}\right),
\end{equation}
whereby $x$ is position along the cross-section slice, $A$ and $x_{p}$ are the amplitude and location of the peak, and $W$ is the Gaussian RMS width. This can be related to the FWHM by: $\mathrm{FWHM}=2\,\sqrt{2\ln{2}}\,W\approx2.35\,W$.

An estimate is made on the number of structures, $N$ that could be present within the intensity profile along with their approximate location, width, and amplitude. Summing the $Y_G$ values for $N$ number of Gaussian curves at each pixel yields the model fit:
\begin{equation}\label{eq:fit}
f\left(x\right) = \sum_{i=1}^N Y_{G\left(i\right)}\left(x\right).
\end{equation}

The closeness of the fit at each pixel, $\chi^2\left(x\right)$ is then determined by measuring the deviation of the fit from the original intensity:
\begin{equation}\label{eq:acc}
\chi^2\left(x\right) = \left(\frac{f\left(x\right)-y\left(x\right)}{\sigma\left(x\right)}\right)^2,
\end{equation}
where $y\left(x\right)$ and $\sigma\left(x\right)$ are the observed \hic\ intensity and Poisson error at each pixel. The overall closeness of fit is then taken as $\sum \chi^2\left(x\right)$, which is then reduced to its smallest value by simultaneously adjusting the free parameters $A$, $x_p$, and $W$ for the $N$ Gaussian curves in $f\left(x\right)$. The minimisation of $\sum \chi^2\left(x\right)$ is performed by using the non-linear least-squares curve fitting method, \textit{MPFIT}\footnote{\textit{MPFIT} is freely-available at: http://purl.com/net/mpfit} \citep{mpfit} which is based on the MINPACK-1 FORTRAN library \citep{minpack}. The one-$\sigma$ uncertainties returned from \textit{MPFIT} are only accurate if the shape of the likelihood surface is well approximated by a parabolic function. Whether fitting multiple Gaussian profiles to each slice satisfies this condition or not would require analysis beyond the scope of this study, however, the one-$\sigma$ uncertainties do provide a lower-bound of the FWHM errors.

To determine the appropriate number of Gaussian profiles, $N$, within a given slice, the AIC model selection is employed. This is done by firstly generating several candidate models, where the number of Gaussian curves differs in each model. The non-linear least-squares curve fitting method is then employed for each candidate model and finally the AICc is then computed. The model with the smallest AICc value is then selected as the preferred model for that \hic\ slice. Once the number of Gaussian profiles contained within a \hic\ slice is determined, the strand width(s) are taken as the Gaussian FWHM value(s).

\citet{klimchuk20} states that this type of analysis is sensitive to background subtraction, and as such we find the number of Gaussian profiles fitted to a \hic\ slice can be influenced by unwanted features such as other loops crossing in-front/behind the loop(s) or underlying moss regions. Whilst concerted efforts are made to avoid these regions, this is not always possible, and these unwanted ``overlaps'' of coronal structures result in differing numbers of sub-element strands being fitted by the AICc method compared to slices either side of the ``overlaps''. Consequently, as this study aims to measure the changes in structural width and emission measure not only along loop segments but also along sub-element strands that may reside within these structures, a consistent number of Gaussian profiles is required to be fitted for the analysed portion of the loop(s). It is reasonable to assume the number of sub-element strands along a loop segment would be consistent and thus the cross-sectional slices where the number of Gaussian profiles determined by the AICc model selection do not match the modal value are likely to be contaminated with other coronal features along the line of sight. With this in mind, the sub-element strands are only analysed for the cross-sectional slices where the number of Gaussian profiles matches the modal value of Gaussian's for the whole loop segment in question. A consequence of this conservative approach to fitting sub-element strands is that it is not always possible to analyse the structures along the entire loop segment. However, measuring the width of the loop-envelope has no such constraints, and the method employed in \citet{williams20a} is adopted with the improved background subtraction discussed in \citet{williams20b}.

\section{Results \& Analysis}\label{sec:res}
A total of eight loops are selected from the \hic\ field-of-view (FOV, Figure\,\ref{fig:fov}) that are relatively free from emission of surrounding coronal structures. From the selection of \hic\ loops chosen for this study, loops A, B, C, and E are from open fan regions\footnote{Visual inspection of the AIA and EUVI full disc images support the expectation that these loops are fan structures and not large, closed loops.} whilst loops D, F, G, and H are magnetically closed structures. In the analysis that follows, the orientation of loops D, F, and G are north-to-south (i.e. position 0 is at the top of the close-up images of Figure\,\ref{fig:fov}) whereas the other loops are orientated south-to-north such that the segments of the structures are traced away from their nearest footpoint. The AICc determined modal number of sub-element strands within loops A, D, E, G, and H is three, whilst loops B and C contain four strands, and Loop F has only two.

As can be seen in the FOV image, selecting a segment of a coronal loop that is isolated from surrounding coronal structures and the ghosting of the CCD mesh cells is a non-trivial task due to the abundance of loop-like structures and the underlying moss in AR\,12712. Focusing on the close-up images of loops A - H in Figure\,\ref{fig:fov} reveals that even the best-case loop segments we have selected here are not completely isolated from other coronal features or the mesh cell ghosting. For example, it is not possible to examine a longer portion of loop A as the segment analysed (denoted by the dashed line) is enclosed on both ends by the ghosting of a mesh cell. Similarly, despite loop D having a high contrast compared to the surrounding corona in the Multi-scale Gaussian Normalisation-enhanced images \citep{morgandruckmuller14}, there are several structures that intersect the loop and thus affect the background subtraction.

\begin{figure*}
\centerline{\includegraphics[width=0.85\textwidth]{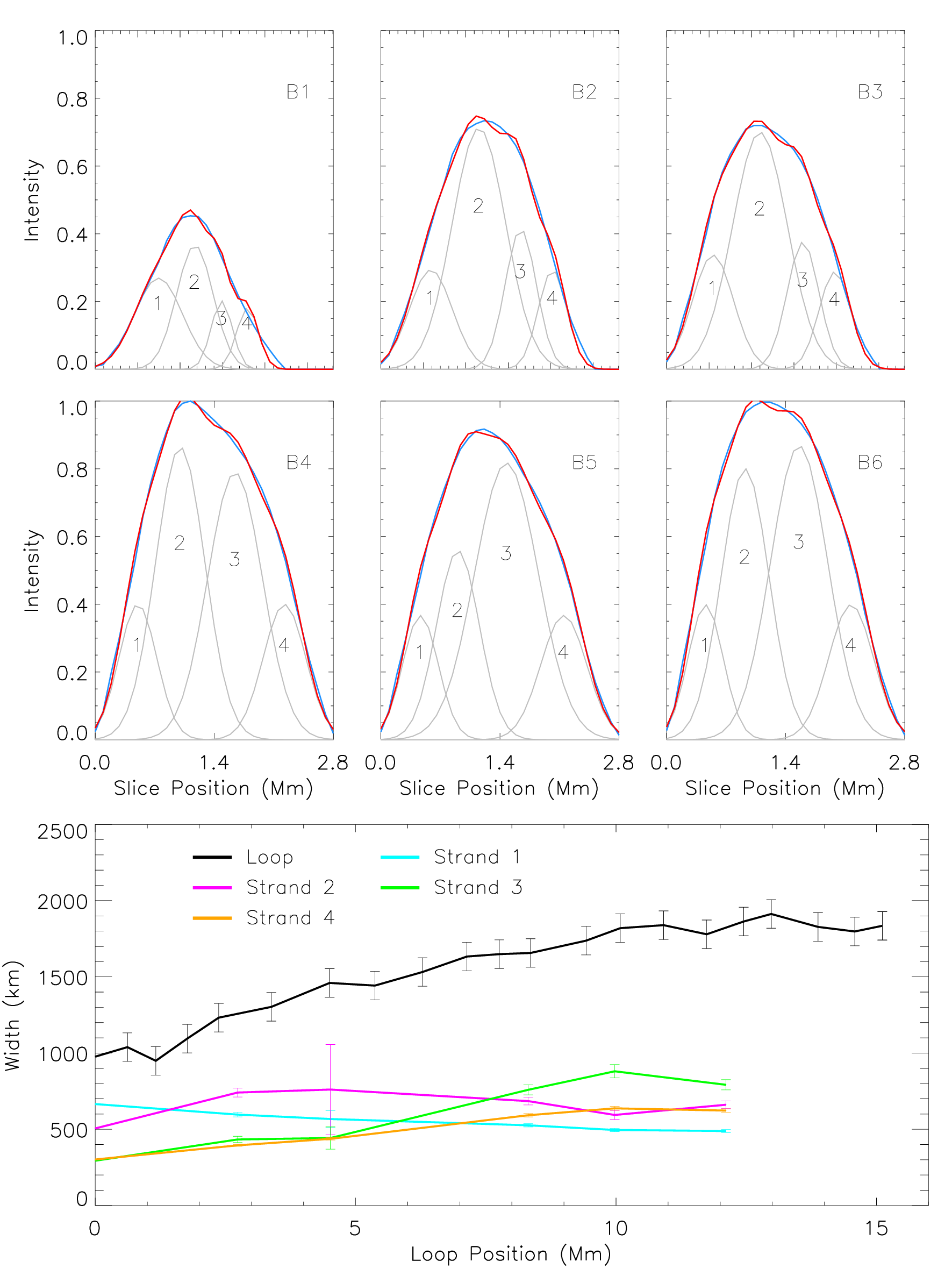}}
\caption{Panels B1\,--\,B6 show the normalised \hic\ intensities (\textit{blue}), the AICc determined fit (\textit{red}), and the Gaussian profiles (\textit{grey}) generated to produce the fit for loop B. Emission is normalised with respect to the maximum peak intensity of panels B1\,--\,B6. The Gaussian profiles are numbered in accordance with the bottom panel which shows the loop (\textit{black}) and strand (\textit{cyan}, \textit{magenta}, \textit{green}, and \textit{orange}) widths as a function of distance along the segment of the loop analysed. The error bars for the loop FWHM are $\pm$1\,\hic\ pixel whereas the strand FWHM error bars are 5$\times$ the one-$\sigma$ uncertainty returned from the curve fitting method. Note that Loop Position 0\,--\,15\,Mm corresponds to south-to-north orientation in Figure\,\ref{fig:fov}.}
\label{fig:loopb}
\end{figure*}
\begin{figure*}
\centerline{\includegraphics[width=0.85\textwidth]{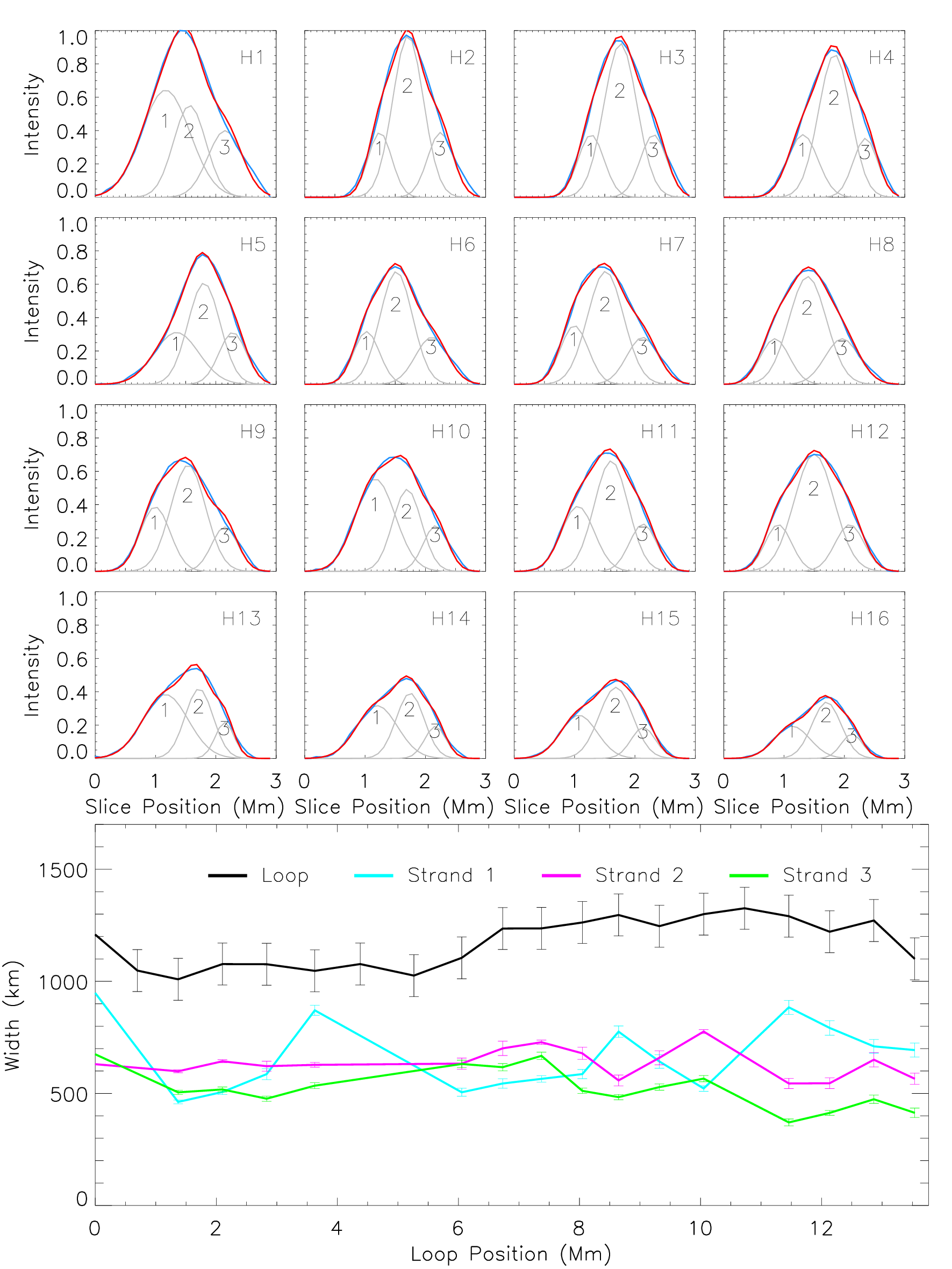}}
\caption{Panels H1\,--\,H16 show the normalised \hic\ intensities (\textit{blue}), the AICc determined fit (\textit{red}), and the Gaussian profiles (\textit{grey}) generated to produce the fit for loop H. Emission is normalised with respect to the maximum peak intensity of panels H1\,--\,H16. The Gaussian profiles are numbered in accordance with the bottom panel which shows the loop (\textit{black}) and strand (\textit{cyan}, \textit{magenta}, and \textit{green}) widths as a function of distance along the segment of the loop analysed. The error bars for the loop FWHM are $\pm$1 \hic\ pixel whereas the strand FWHM error bars are 5$\times$ the one-$\sigma$ uncertainty returned from the curve fitting method.Note that Loop Position 0\,--\,13\,Mm corresponds to south-to-north orientation in Figure\,\ref{fig:fov}.}
\label{fig:looph}
\end{figure*}
\begin{figure}
\centerline{\includegraphics[width=0.45\textwidth]{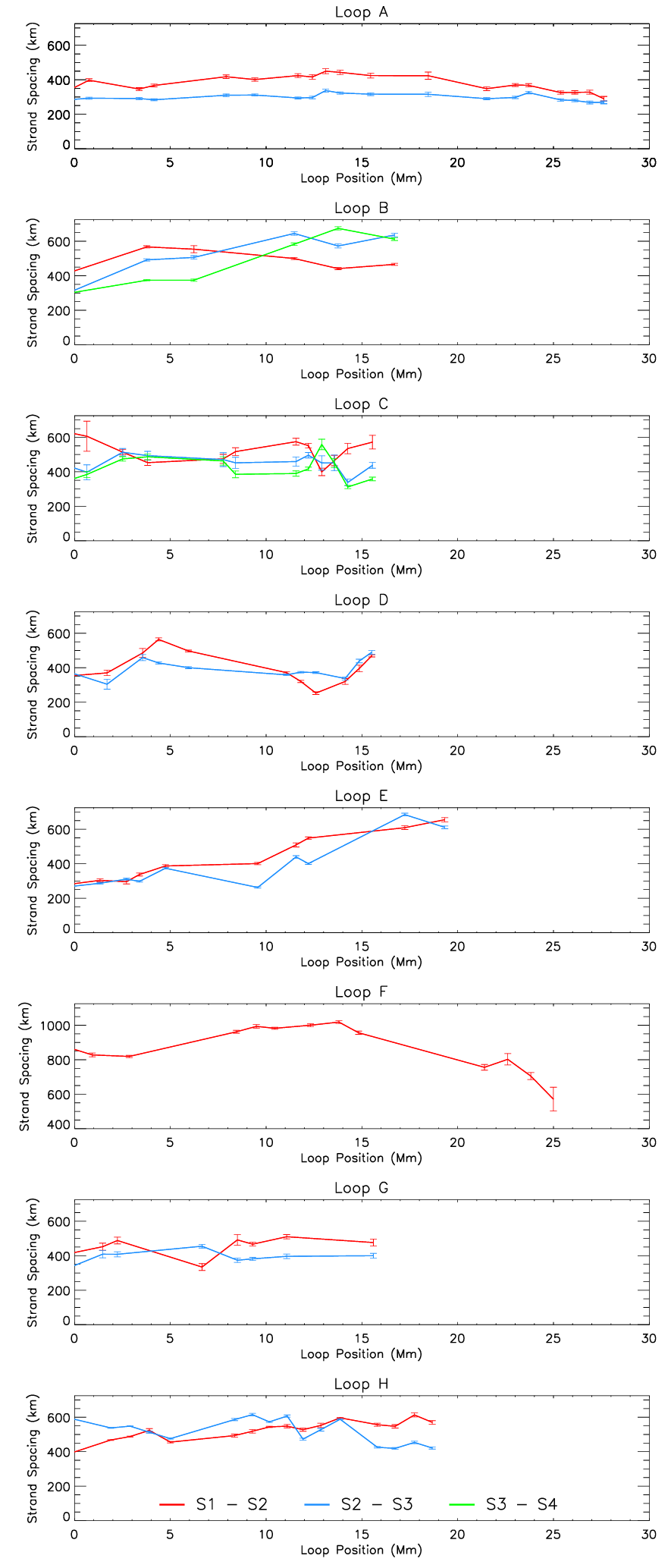}}
\caption{Figure shows the separation between the peak intensity of strands as a function of distance along the loop segments analysed for strands 1 and 2 (\textit{blue}), strands 2 and 3 (\textit{red}), and strands 3 and 4 (\textit{green}) for the eight loops investigated in this study. Note that increasing Loop Position for loops A, B, C, D, and H (E, F, and G) corresponds to south-to-north (north-to-south) orientation in Figure\,\ref{fig:fov}.}
\label{fig:spacing}
\end{figure}

\subsection{Width Analysis}\label{sec:widanal}
In Figures\,\ref{fig:loopb} and \ref{fig:looph} the top panels labelled B1\,--\,B6 and H1\,--\,H16 show the \hic\ cross-sectional intensity (\textit{blue}), the AICc determined best-fit (\textit{red}), and the Gaussian profiles (\textit{grey}) that generate the fit. The intensity of these panels are normalised with respect to the maximum peak intensity of the loop segment being considered. The same figures for the other six loops are shown in the Appendix (Figures\,\ref{fig:loopa}\,--\,\ref{fig:loopg}).  In all these plots for loops A\,--\,H it is possible to follow the individual strands from one cross-sectional slice to the next along the loop segments analysed. As with previous results using this method \citep{williams20b}, the AICc determined fits closely match the observed \hic\ cross-sectional profiles in most instances, with some minor exceptions such as slices A3 and F6. Here, it would be possible to reduce the deviance between the generated fits and the \hic\ emission by adding further Gaussian profiles, however, in doing so the curve fitting model would no longer be supported by the AICc model selection method and would increase the likelihood of over-fitting the data along the length of the loop segments analysed.

The bottom panels of Figures\,\ref{fig:loopb} and \ref{fig:looph} show the variation in loop (\textit{black}) and strand (\textit{cyan, magenta, green, orange}) widths as a function of loop position. The error bars for the loop width indicate a \hic\ pixel width (93.525\,km) whereas the error bars of the strands are the one-$\sigma$ errors returned from \textit{MPFIT}. From the eight loops considered, six of them show no signs of significant expansion, whereas the two loops that do (B and E) are open fan structures and may be expected to expand. Curiously, A and C are also open fan structures yet their widths remain largely constant along the segments analysed. It could be the case that these structures have undergone expansion closer to the footpoints as with loops B and E, however, this cannot be verified in this data-set with our current methods.

Figure\,\ref{fig:spacing} shows the separation between the peak intensity of neighbouring sub-element strands as a function of loop position for the loops A\,--\,H. Increasing loop position for loops A, B, C, D, and H (E, F, and G) correspond to south-to-north (north-to-south) orientation in Figure\,\ref{fig:fov}. Focusing upon the six loops that show no obvious expansion, their sub-element strands also exhibit no expansion (Figures\,\ref{fig:looph},\,\ref{fig:loopa},\,\ref{fig:loopc},\,\ref{fig:loopd},\,\ref{fig:loopf},\,\ref{fig:loopg}) or separation between adjacent strands (Figure\,\ref{fig:spacing}). Of these loops, only loops C and F show initial and final widths that are above the observational error of two \hic\ pixel widths (187.05\,km). For loop C, this level of expansion is insignificant (80\,km above 2 \hic\ pixels), and the close proximity of a loop to the east whose intensity increases when tracing the structure in a south-to-north orientation (Figure\,\ref{fig:fov}) may effect the background subtraction performed on loop C, which would account for the decreased width seen from loop position 14\,Mm onward in Figure\,\ref{fig:loopc}. For loop F, the width is predominantly the same along the loop segment until a sudden decrease is seen in the final three cross-sectional slices (Figure\,\ref{fig:loopf}). As can be seen in the cross-sectional slices and from the visual reference in Figure\,\ref{fig:fov}, the loop intensity decreases relative to the surrounding loop bundle and thus the magnitude of the loop intensity relative to the background emission that is subtracted decreases, which may contribute to the contraction in loop width indicated in Table\,\ref{table:emission}.

To quantify the total expansion ($\Gamma_{\mathrm{tot}}$) of each loop/strand, the final and initial widths of the segment analysed are calculated as the ratio 
\begin{equation*}
    \Gamma_{\mathrm{tot}} = \frac{W_{\mathrm{{final}}}}{W_{\mathrm{initial}}}.
\end{equation*}
The expansion rate ($\Gamma_{\mathrm{rate}}$) is then simply expressed as
\begin{equation*}
    \Gamma_{\mathrm{rate}}=\frac{\Gamma_{\mathrm{tot}}}{l},
\end{equation*}
where $l$ is the length of the loop/strand segment.

Employing this on loops B and E, it can be seen that they expand a total of $\approx860$\,km and $\approx980$\,km, giving expansion rates ($\Gamma_{\mathrm{rate}}$) along their lengths of 4.2\,\% and 6.8\,\% per Mm, respectively (Table\,\ref{table:emission}). For Loop B, the widths of strands 1 and 2 remain within two pixel widths and as such do not contribute to the expansion of Loop B, whilst strands 3 and 4 expand $\approx500$\,km and $\approx320$\,km. Furthermore, Figure\,\ref{fig:spacing} reveals that whilst the separation between strands 1 and 2 remains roughly constant, strands 2 and 3, and strands 3 and 4 spread apart along the loop segment under consideration. If we consider the findings of \citet{mcclymont94} that a twisted structure exhibits less expansion than an untwisted structure, then it may be that strands 1 and 2 are twisted structures whilst strands 3 and 4 are subjected to less twist and thus expand with height.

For loop E, all three strands show expansion above two pixel widths ($\approx260$\,km, $\approx460$\,km, and $\approx220$\,km, respectively) and the separation between all the sub-element strands also increases along the loop segment analysed. Thus, the expansion observed in the open fan structures B and E are the result of sub-element strands both expanding and separating from each other simultaneously.

\begin{table*}
\centering
\caption{Pearson coefficients and expansion factors for all the loops/strands analysed.}
\begin{tabular}{ |c|c|c|c|c|c|c| }
\hline
Structure & Loop & Pearson & Initial & Final & Total & Expansion \\
& Type & Coefficient & Width (km) & Width (km) &  Expansion & Rate (Mm$^{-1}$) \\
\hline
Loop A & Open Fan & 0.198 & 899.3 & 754.9 & -0.161 & -0.005 \\
Strand 1 && -0.247 & 437.3 & 495.4 & 0.133 & 0.005 \\
Strand 2 && 0.715 & 375.5 & 281.8 & -0.249 & -0.008 \\
Strand 3 && -0.587 & 327.9 & 321.4 & -0.020 & -0.001 \\
\hline
Loop B & Open Fan & 0.995 & 976.5 & 1835.0 & 0.879 & 0.042 \\
Strand 1 && -0.926 & 665.3 & 488.5 & -0.266 & -0.021 \\
Strand 2 && 0.764 & 505.7 & 661.0 & 0.307 & 0.021 \\
Strand 3 && 0.981 & 293.8 & 791.9 & 1.696 & 0.102 \\
Strand 4 && 0.949 & 301.9 & 622.8 & 1.063 & 0.084 \\
\hline
Loop C & Open Fan & 0.772 & 1581.4 & 1314.1 & -0.169 & -0.169 \\
Strand 1 && 0.931 & 791.2 & 673.9 & -0.148 & -0.011 \\
Strand 2 && 0.733 & 790.5 & 690.2 & -0.127 & -0.008 \\
Strand 3 && 0.911 & 470.6 & 404.6 & -0.140 & -0.009 \\
Strand 4 && 0.538 & 490.5 & 387.7 & -0.210 & -0.014 \\
\hline
Loop D & Closed & 0.314 & 784.7 & 768.6 & -0.021 & -0.001 \\
Strand 1 && 0.525 & 768.9 & 403.6 & -0.475 & -0.031 \\
Strand 2 && 0.105 & 447.2 & 568.4 & 0.271 & 0.018 \\
Strand 3 && 0.465 & 442.6 & 493.2 & 0.115 & 0.007 \\
\hline
Loop E & Open Fan & 0.869 & 651.9 & 1631.1 & 1.502 & 0.068 \\
Strand 1 && 0.844 & 510.5 & 768.3 & 0.505 & 0.027 \\
Strand 2 && 0.947 & 286.3 & 749.7 & 1.618 & 0.073 \\
Strand 3 && 0.789 & 297.1 & 513.6 & 0.729 & 0.038 \\
\hline
Loop F & Closed & 0.384 & 1664.0 & 1199.8 & -0.279 & -0.010 \\
Strand 1 && 0.016 & 1096.0 & 1091.1 & -0.004 & $<$-0.001 \\
Strand 2 && 0.464 & 745.3 & 801.5 & 0.075 & 0.003 \\
\hline
Loop G & Closed & 0.312 & 1054.4 & 1096.2 & 0.040 & 0.003 \\
Strand 1 && 0.143 & 694.8 & 651.1 & -0.063 & -0.004 \\
Strand 2 && 0.490 & 434.7 & 447.9 & 0.030 & 0.002 \\
Strand 3 && -0.053 & 429.4 & 359.7 & -0.162 & -0.010 \\
\hline
Loop H & Closed & -0.437 & 1208.5 & 1100.5 & 0.052 & 0.003 \\
Strand 1 && 0.388 & 947.7 & 693.3 & -0.268 & -0.014 \\
Strand 2 && 0.395 & 630.5 & 565.5 & -0.103 & -0.006 \\
Strand 3 && 0.520 & 674.6 & 413.6 & -0.387 & -0.207 \\
\hline
\end{tabular}
\label{table:emission}
\end{table*}


\subsection{Intensity vs. Width}\label{sec:emission}
\begin{figure}
\centerline{\includegraphics[width=0.5\textwidth]{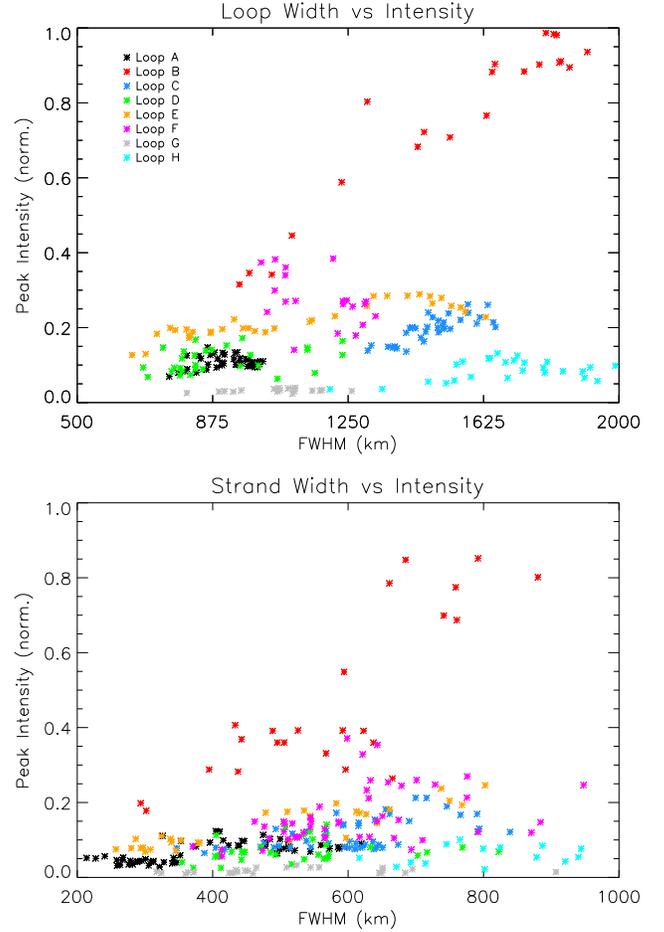}}
\caption{\textit{Top}: Peak intensities are plotted against their widths for loops A\,--\,H. \textit{Bottom}: Peak intensities of the strands contained within loops A\,--\,H are plotted against their widths. No distinction is made here about individual strands but rather the strands as a collective for each loop, i.e. strands 1\,--\,3 of loop A are all plotted as \textit{black} asterisks, strands 1\,--\,3 of loop B are plotted as \textit{red} asterisks, strands 1\,--\,4 of loop C are plotted as \textit{blue} asterisks, etc.}
\label{fig:ivw}
\end{figure}
Following the work of \citet{klimchuk20}, this section investigates the relationship between the peak intensity and width of the selected loops and, for the first time, extends this analysis to their sub-element strands. As is discussed in \citet{klimchuk20}, a negative correlation between width and intensity may indicate that a loop has an elliptical cross-section whereas if there is no or positive correlation, the cross-section is likely to be roughly circular.

Figure\,\ref{fig:ivw} top (bottom) shows the peak intensity of the loops (strands) as a function of FWHM. From visual inspection it can be seen that the peak intensities of loops B, C, and E appear strongly correlated to the widths of the structures whilst the remaining loops have weaker correlations where it is not possible to accurately determine whether the structures are positively or negatively correlated. As for the peak intensity of all the strands, they appear to predominantly be positively correlated to their widths -- albeit to varying degrees with the strands of loop B showcasing the strongest correlation. 

To quantify whether the peak intensity of cross-sectional slices along a segment of a loop/strand is positively or negatively correlated to the FWHM, the Pearson correlation coefficient is determined, the results of which are presented in Table\,\ref{table:emission}. The Pearson correlation coefficient, $P$ is defined as
\begin{equation}
    P = \frac{\mathrm{cov}\left(\mathrm{FWHM},I\right)}{\sigma_\mathrm{FWHM}\,\sigma_I},
\end{equation}
where $\mathrm{cov}\left(W,I\right)$ is the covariance of FWHM and intensity, whilst the denominator is the product of their standard deviations. If a structure has a positive (negative) Pearson correlation then as the width increases the intensity increases (decreases). The coefficient magnitude varies between -1 and 1 with values of $P \approx 0$ corresponding to uncorrelated data, whilst values of -1 and 1 correspond to perfectly correlated data.

From Table\,\ref{table:emission}, it is found that all the peak intensities of the four open fan loops are positively correlated to their width, with loops B, C, and E being strongly correlated whilst loop A displays weak-to-moderate positive correlation. The weaker correlation of loop A is likely due to the fact that the structure exhibits minimal variation in the peak intensity and its width along the segment analysed, and so deducing a relationship between the two is more difficult. As for the sub-element strands within the four open fan structures, a total of three strands are negatively correlated (loop A, strands 1, 3 and loop B, strand 1), with the remaining eleven all exhibiting strong positive Pearson correlation coefficients. This indicates that as the structures broaden, peak intensities increase for open fan loops and their sub-element strands.

As for the magnetically closed structures, three of the four loops (D, F, and G) are positively correlated whilst loop H is negatively correlated. All four loops have modest Pearson correlation coefficient magnitudes that are weaker than all the open fan loops bar loop A. Peculiarly, all the Pearson coefficients bar one of the sub-element strands are positively correlated for the magnetically closed structures despite the fact that loop H is negatively correlated. The other strand (loop G, strand 3), whilst having a negative Pearson coefficient, the absolute magnitude is such that the strand exhibits no linear correlation and thus is considered to be uncorrelated.

Focusing on loop H, the loop width broadens as one travels along loop position (Figure\,\ref{fig:looph} \textit{bottom}) whilst the peak intensity of the cross-sectional slices decreases (Figure\,\ref{fig:looph} H1\,--\,H16). Thus, there should be a negative correlation between the intensity and width for the loop. If we now consider the sub-element strands, such as strand 2 say, it can be seen that as the peak intensity fluctuates, the FWHM remains relatively constant along the analysed segment whilst the spacing between strands 1 and 2 (2 and 3) increases (decreases) (Figure\,\ref{fig:spacing}) as the loop expands, i.e. the strands spread out (overlap more) along the integrated line-of-sight. All these factors together may explain a weak positive Pearson correlation coefficient for the strands of loop H despite the loop being negatively correlated.

As is noted in \citet{klimchuk20}, if a loop or strand were to be non-circular (such as an oval) and twisted, there would be an inverse relationship between the intensity and width. That is, the loop would have a higher (lower) peak intensity at narrower (broader) loop widths if relatively constant density and temperature are assumed. The majority of our results for open fan and closed loops, along with \citet{klimchuk20}, and nine of the thirteen structures analysed by \citet{mccarthy21} indicate that the relationship between intensity and width may be positively correlated and thus their cross sections are approximately circular. However, it is worth noting that one magnetically closed loop, and three open fan strands of this study along with four loops analysed by \citet{mccarthy21} do support the idea that some loops and sub-element strands may be non-circular.
\section{Concluding Remarks}\label{sec:conc}
Employing the analysis methods of \citet{williams20a,williams20b}, this article investigates the relationship between width and peak intensity of eight relatively isolated coronal loops and extends this for the first time to sub-element strands within the \hic\ data. These loops and strands are traced along segments 10\,--\,30\,Mm in length, which are of comparable length to the loops analysed with the data from the first Hi-C mission \citep{klimchuk20} but are some 20\,--\,50\,Mm shorter than those in the recent study by \citet{mccarthy21}.

The population of the eight loops analysed is evenly split between open fan (loops A, B, C, and E) and magnetically closed (loops D, F, G, and H) structures and are thus treated independently from each other in our width expansion analysis. The segments analysed for loops B and E reside near the visible footpoints (Figure\,\ref{fig:fov}) of the open fan structures and thus undergo significant expansion whereby their widths increase an additional 80\,\% and 150\,\% (see Total Expansion, Table\,\ref{table:emission}) of their initial value. Conversely, the segments analysed for loops A and C (Figures\,\ref{fig:loopa} and \ref{fig:loopc}) show minimal width variation along their length. These results suggest that open fan loops may expand rapidly near their footpoints and after which their cross-sectional width may remain relatively constant. With that said, a more comprehensive data-set is required from future high resolution missions to confirm this as current and archival data lack the resolving power to accurately quantify the expansion seen with \hic.

As with previous loop width studies \citep{klimchuk92,klimchuk00,klimchuk15,fuentes06}, the magnetically closed structures analysed in this article exhibit little expansion with three of the four loops showing $<6$\,\% (108\,km; $<1.2$\,\hic\ pixels) expansion (Table\,\ref{table:emission}) along the analysed segments. Similarly, the width and separation distance of the sub-element strands of these loops remain largely constant along the loop portions under consideration in this study. One caveat to this is that loop D can be seen to expand and contract periodically. 

As discussed by \citet{klimchuk20}, the largest error in coronal loop width analysis is a direct result of performing background subtraction due to the relative magnitudes of the loop and line-of-sight emission that is removed. Close inspection of the FOV image (Figure\,\ref{fig:fov}) reveals multiple structures intersecting/crossing loop D, which makes it more difficult to employ a consistent background removal along the loop segment. As such, these variations in loop width are likely to be the caused by background subtraction due to the crossing of multiple structures along the line-of-sight rather than because the loop is a twisted structure of non-circular cross-section \citep{malanushenko13}. 

Generalising the data-set as a whole, the expansion of loops can be quantified by the sub-element strands broadening and separating from each other simultaneously along the segments considered which leads to the increasing cross-sectional loop widths seen in loops B and E. Similarly, the lack of expansion seen in the other six structures is caused by strands whose widths and separation remain relatively constant along the loop segments analysed.

Upon exploring the relationship between loop width and peak intensity, this article finds that seven of the eight loops analysed ($\approx88$\,\%) are positively correlated, with the other loop being negatively correlated. These results match previous findings \citep{west14,kucera19,klimchuk20} with \citet{klimchuk20} being a notable comparison due to their work also examining loops of similar lengths to the ones in this study with Hi-C, albeit with the 193\,\AA\ data-set. As such, our results support the \citet{klimchuk20} hypothesis that loops are more likely to exhibit circular cross sections rather than elliptical ones \citep{malanushenko13,mccarthy21}, which would be the case if the peak intensities and widths of the cross sectional slices were negatively correlated. As for the coronal strands, we obtain a similar outcome with 20 strands (80\,\%) showing positive correlation, 3 strands (12\,\%) showing negative correlation and 2 strands (8\,\%) being uncorrelated.

The findings presented in this article for magnetically closed coronal loops and their sub-element strands support the discussion of \citet{klimchuk00} and \citet{williams20a,williams20b} in regards to coronal heating. That is, whatever mechanism is responsible for the coronal heating, theoretical models may need to reproduce monolithic strands whose cross-sections (i) have widths at the same order of an AIA pixel ($\approx435$\,km), and (ii) are approximately circular along their length. These implications for coronal heating are contingent on coronal loops being compact structures that are subjected to twist. In particular, fitting multiple nearly-cylindrical Gaussian-shaped strands relies heavily on the assumption that curvilinear features in the \hic\ image plane do indeed correspond to compact structures in three dimensions. Whilst this assumption is plausible, other explanations may be equally valid in explaining the cross-sectional profiles observed. As is discussed in \citet{mccarthy21}, monolithic loops when viewed from a single vantage point -- such as AIA or in this study \hic\ -- may in fact be loops with non-fixed aspect ratios and/or exhibit anisotropic expansion. They may also be the result of  projection effects of complicated bright manifolds in the corona, or some other complex density enhancement. Thus, further three dimensional analysis is required to confirm whether coronal loops are possibly well represented by a collection of sub-element strands.

As with the \citet{west14} (11 loops), \citet{klimchuk20} (20 loops) and \citet{mccarthy21} (13 loops) studies, this analysis is on a small sample size -- largely due to the difficulty in tracing isolated segments of coronal loops within the \hic\ data -- and as suggested by \citet{malanushenko13} may be subject to selection bias. Further high resolution data is required, such as from \textit{Solar Orbiter}'s Extreme Ultraviolet Imager (EUI; $\approx 0.1$\arc\,pixel\textsuperscript{-1} at 0.28\,AU) instrument, to explore the relationship between intensity and width in the hope of understanding the fundamental geometry of these important structures in the quest for solving the coronal heating problem.

\begin{acknowledgements}
TW and HM gratefully acknowledge support by Leverhulme grant RPG-2019-361. HM also acknowledges STFC grant ST/S000518/1.
\end{acknowledgements}

\begin{appendix}
\section{Fitting Coronal Loop Sub-Elements}\label{app}
\begin{figure*}
\centerline{\includegraphics[width=0.85\textwidth]{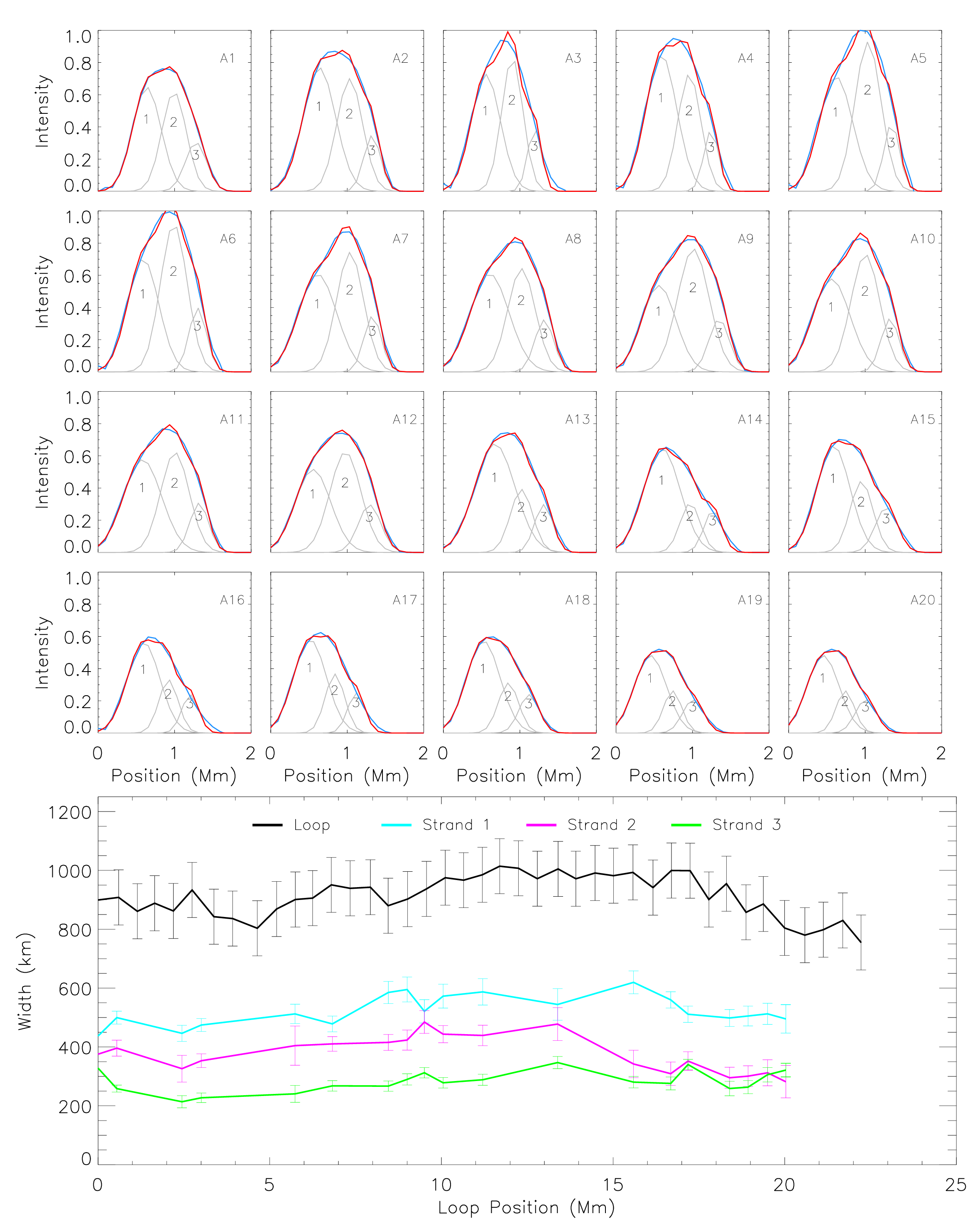}}
\caption{Panels A1\,--\,A20 show the normalised \hic\ intensities (\textit{blue}), the AICc determined fit (\textit{red}), and the Gaussian profiles (\textit{grey}) generated to produce the fit for loop A. Emission is normalised with respect to the maximum peak intensity of panels A1\,--\,A20. The Gaussian profiles are numbered in accordance with the bottom panel which shows the loop (\textit{black}) and strand (\textit{cyan}, \textit{magenta}, and \textit{green}) widths as a function of distance along the segment of the loop analysed. The error bars for the loop FWHM are $\pm$1\,\hic\ pixel whereas the strand FWHM error bars are 5$\times$ the one-$\sigma$ uncertainty returned from the curve fitting method.Note that Loop Position 0\,--\,25\,Mm corresponds to south-to-north orientation in Figure\,\ref{fig:fov}.}
\label{fig:loopa}
\end{figure*}

\begin{figure*}
\centerline{\includegraphics[width=0.85\textwidth]{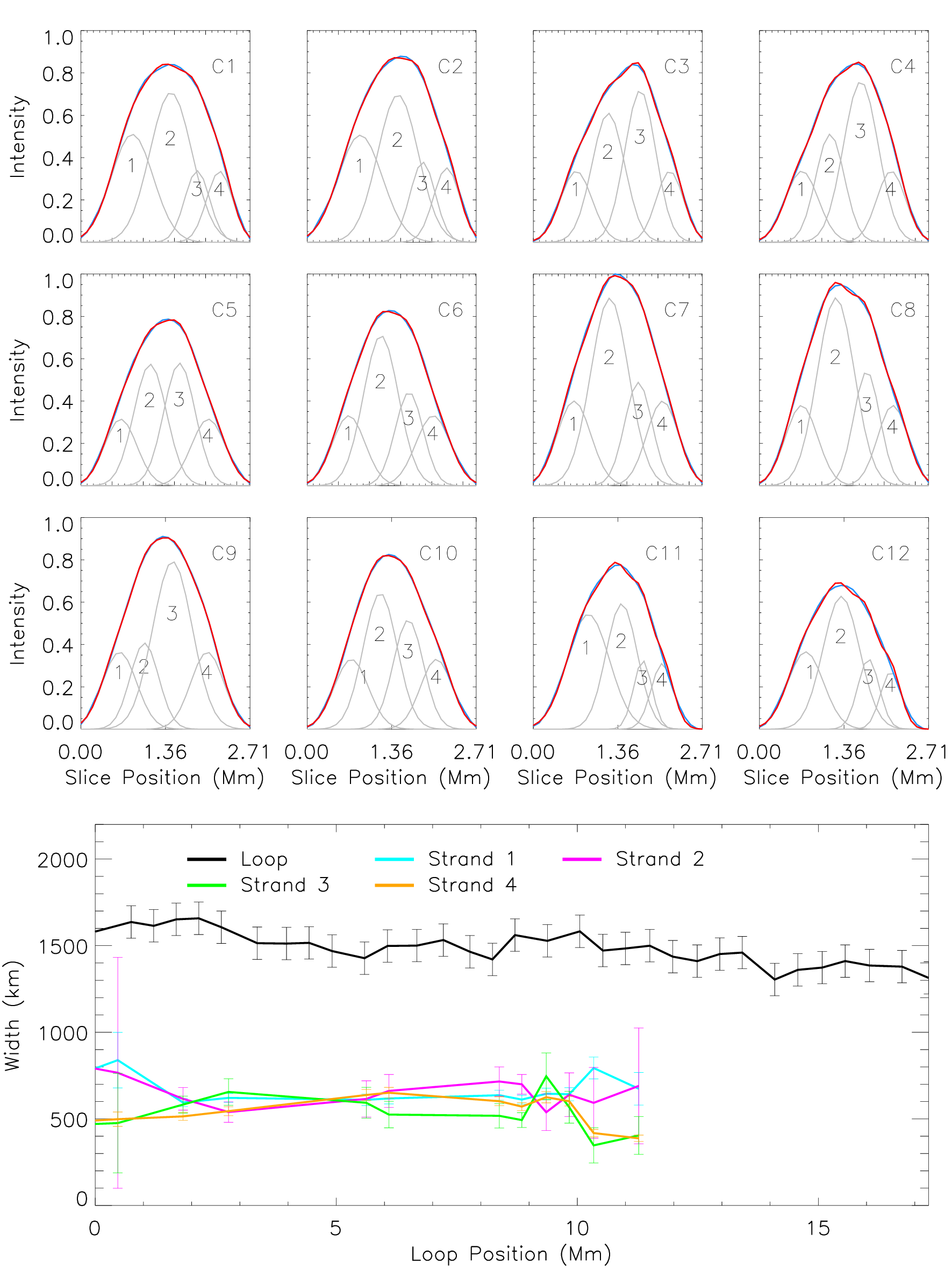}}
\caption{Panels C1\,--\,C12 show the normalised \hic\ intensities (\textit{blue}), the AICc determined fit (\textit{red}), and the Gaussian profiles (\textit{grey}) generated to produce the fit for loop C. Emission is normalised with respect to the maximum peak intensity of panels C1\,--\,C12. The Gaussian profiles are numbered in accordance with the bottom panel which shows the loop (\textit{black}) and strand (\textit{cyan}, \textit{magenta}, \textit{green}, and \textit{orange}) widths as a function of distance along the segment of the loop analysed. The error bars for the loop FWHM are $\pm$1\,\hic\ pixel whereas the strand FWHM error bars are 5$\times$ the one-$\sigma$ uncertainty returned from the curve fitting method. Note that Loop Position 0\,--\,17\,Mm corresponds to south-to-north orientation in Figure\,\ref{fig:fov}.}
\label{fig:loopc}
\end{figure*}

\begin{figure*}
\centerline{\includegraphics[width=0.85\textwidth]{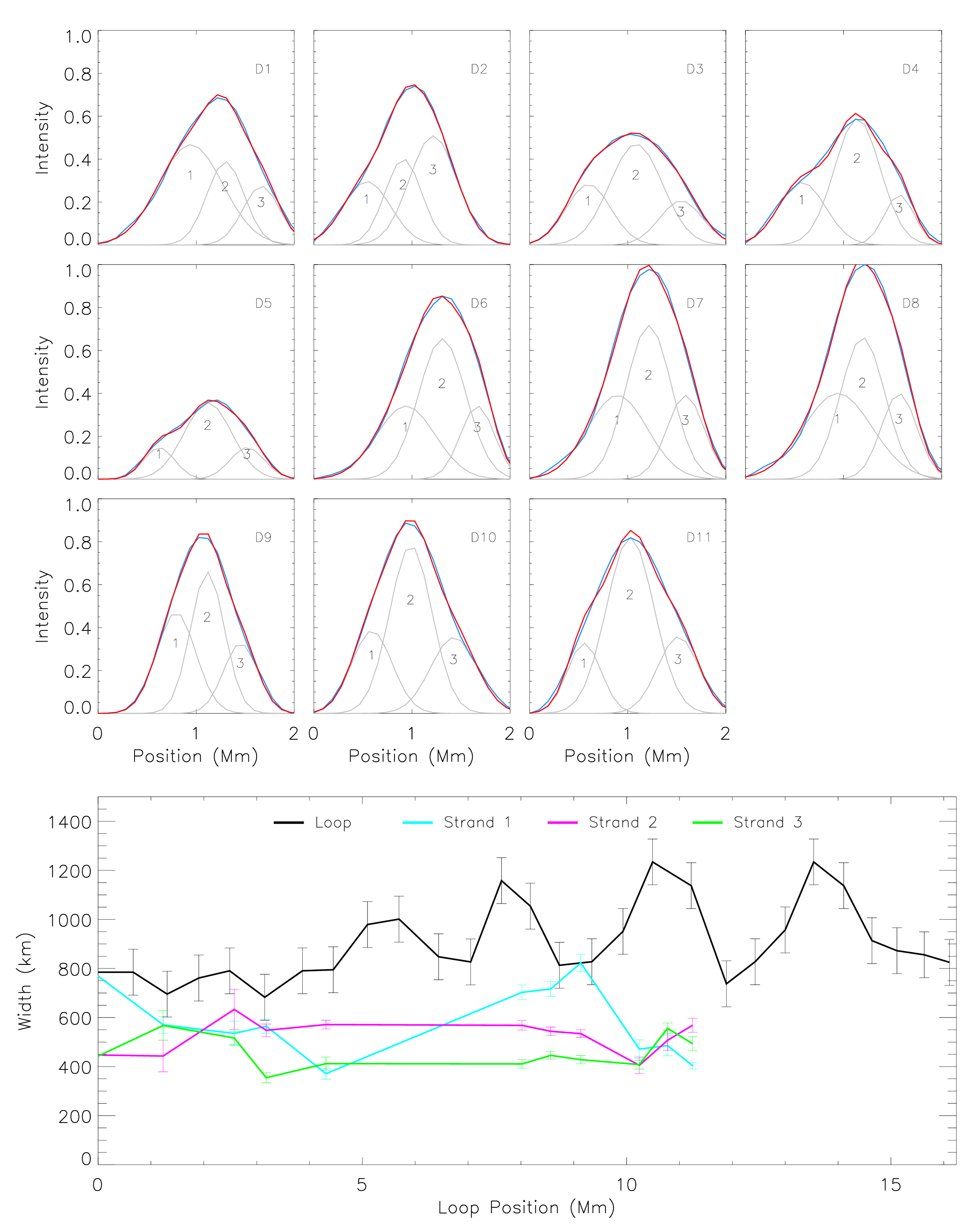}}
\caption{Panels D1\,--\,D11 show the normalised \hic\ intensities (\textit{blue}), the AICc determined fit (\textit{red}), and the Gaussian profiles (\textit{grey}) generated to produce the fit for loop D. Emission is normalised with respect to the maximum peak intensity of panels D1\,--\,D11. The Gaussian profiles are numbered in accordance with the bottom panel which shows the loop (\textit{black}) and strand (\textit{cyan}, \textit{magenta}, and \textit{green}) widths as a function of distance along the segment of the loop analysed. The error bars for the loop FWHM are $\pm$1\,\hic\ pixel whereas the strand FWHM error bars are 5$\times$ the one-$\sigma$ uncertainty returned from the curve fitting method. Note that Loop Position 0\,--\,16\,Mm corresponds to south-to-north orientation in Figure\,\ref{fig:fov}.}
\label{fig:loopd}
\end{figure*}

\begin{figure*}
\centerline{\includegraphics[width=0.85\textwidth]{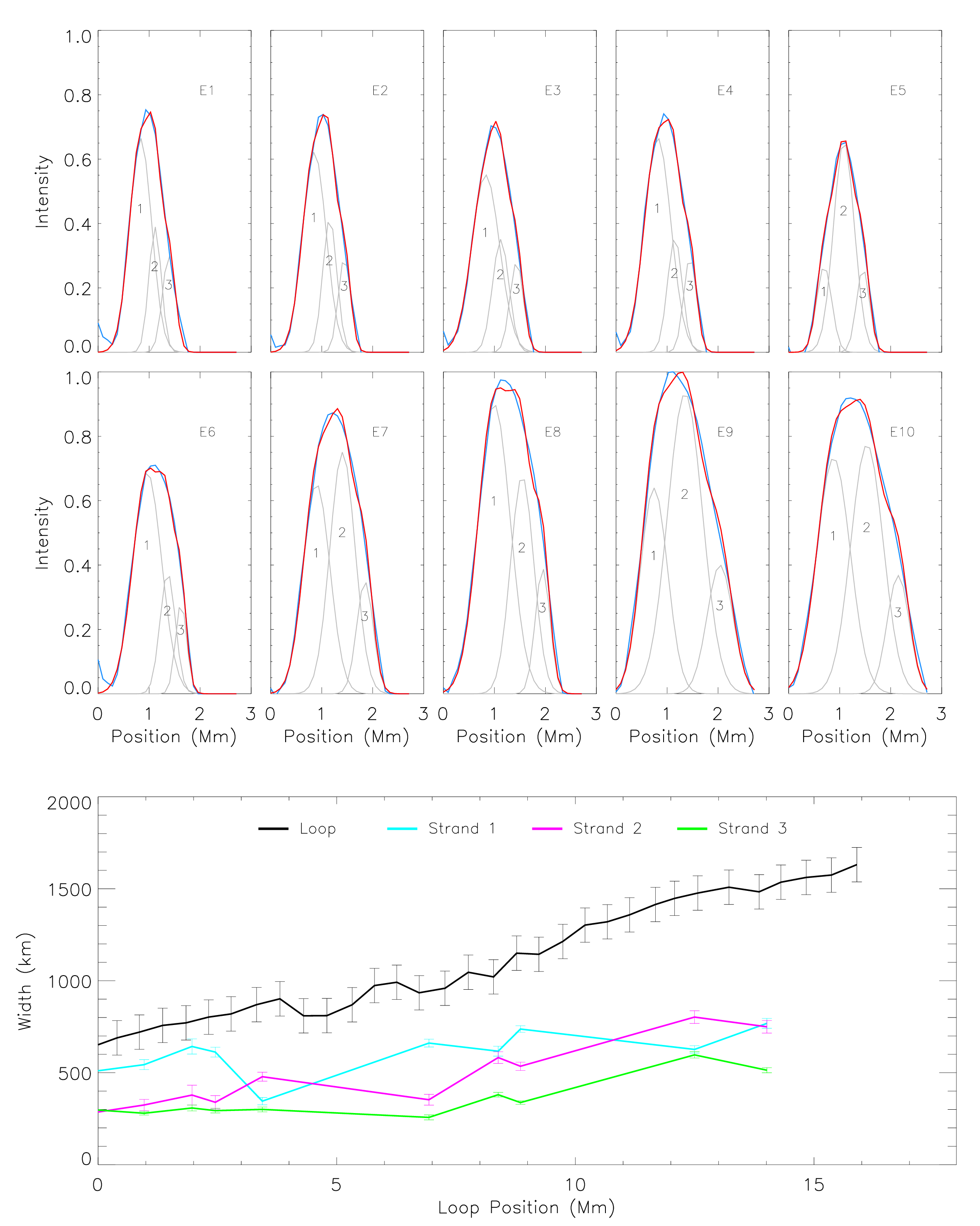}}
\caption{Panels E1\,--\,E10 show the normalised \hic\ intensities (\textit{blue}), the AICc determined fit (\textit{red}), and the Gaussian profiles (\textit{grey}) generated to produce the fit for loop E. Emission is normalised with respect to the maximum peak intensity of panels E1\,--\,E10. The Gaussian profiles are numbered in accordance with the bottom panel which shows the loop (\textit{black}) and strand (\textit{cyan}, \textit{magenta}, and \textit{green}) widths as a function of distance along the segment of the loop analysed. The error bars for the loop FWHM are $\pm$1\,\hic\ pixel whereas the strand FWHM error bars are 5$\times$ the one-$\sigma$ uncertainty returned from the curve fitting method. Note that Loop Position 0\,--\,18\,Mm corresponds to north-to-south orientation in Figure\,\ref{fig:fov}.}
\label{fig:loope}
\end{figure*}

\begin{figure*}
\centerline{\includegraphics[width=0.85\textwidth]{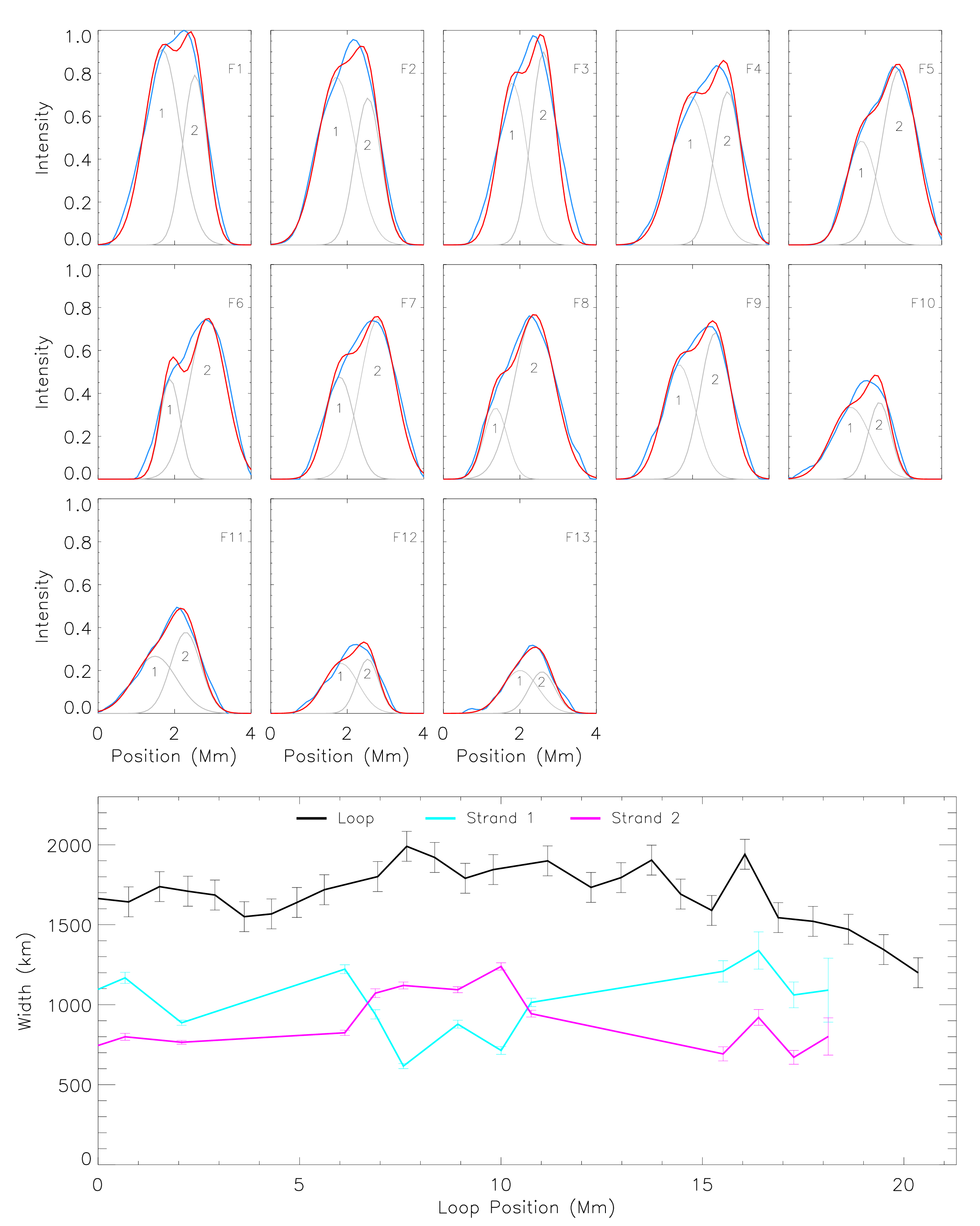}}
\caption{Panels F1\,--\,F13 show the normalised \hic\ intensities (\textit{blue}), the AICc determined fit (\textit{red}), and the Gaussian profiles (\textit{grey}) generated to produce the fit for loop F. Emission is normalised with respect to the maximum peak intensity of panels F1\,--\,F13. The Gaussian profiles are numbered in accordance with the bottom panel which shows the loop (\textit{black}) and strand (\textit{cyan}, \textit{magenta}, and \textit{green}) widths as a function of distance along the segment of the loop analysed. The error bars for the loop FWHM are $\pm$1\,\hic\ pixel whereas the strand FWHM error bars are 5$\times$ the one-$\sigma$ uncertainty returned from the curve fitting method. Note that Loop Position 0\,--\,21\,Mm corresponds to north-to-south orientation in Figure\,\ref{fig:fov}.}
\label{fig:loopf}
\end{figure*}

\begin{figure*}
\centerline{\includegraphics[width=0.85\textwidth]{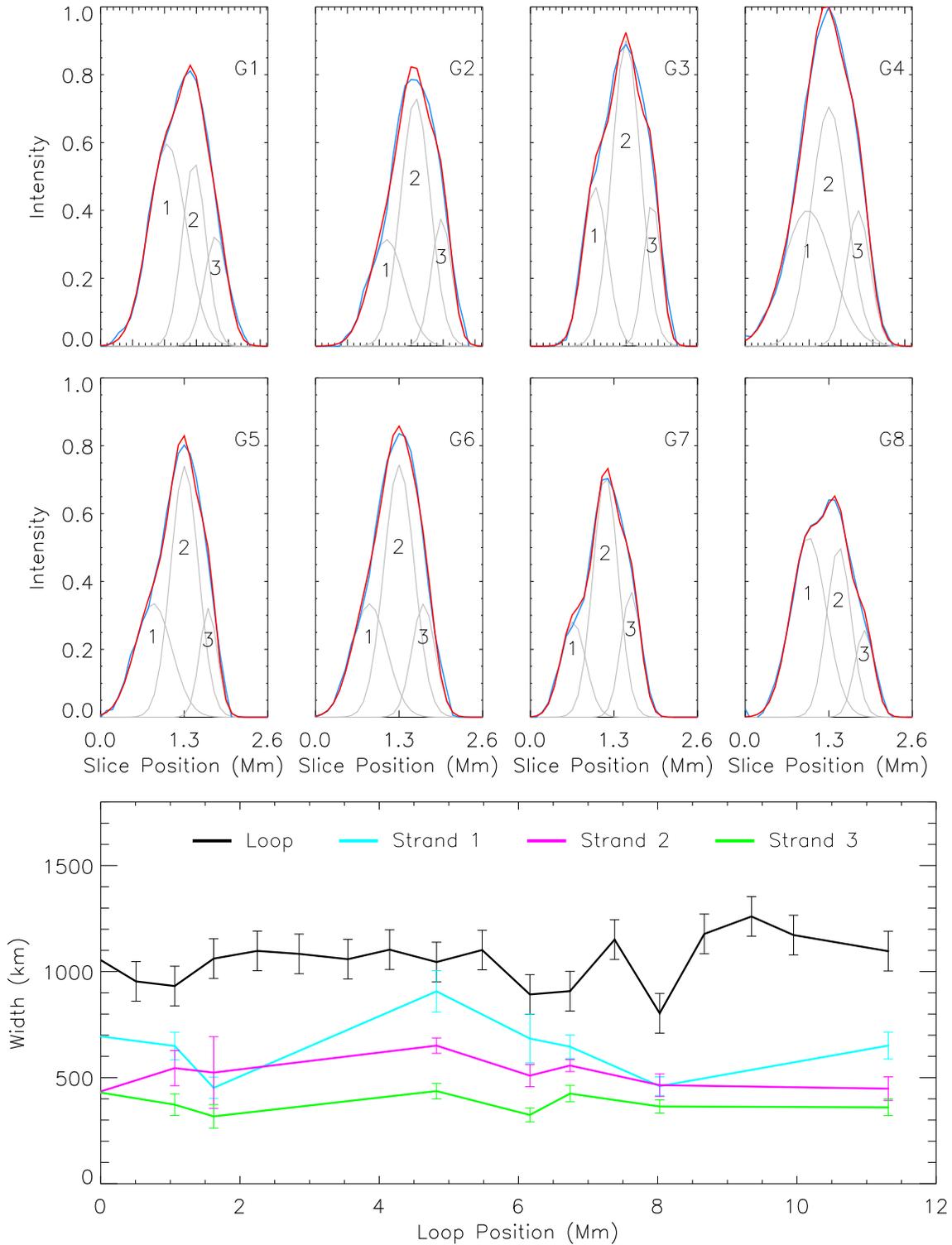}}
\caption{Panels G1\,--\,G11 show the normalised \hic\ intensities (\textit{blue}), the AICc determined fit (\textit{red}), and the Gaussian profiles (\textit{grey}) generated to produce the fit for loop E. The Gaussian profiles are numbered in accordance with the bottom panel which shows the loop (\textit{black}) and strand (\textit{cyan}, \textit{magenta}, and \textit{green}) widths as a function of distance along the segment of the loop analysed. The error bars for the loop FWHM are $\pm$1\,\hic\ pixel whereas the strand FWHM error bars are 5$\times$ the one-$\sigma$ uncertainty returned from the curve fitting method. Note that Loop Position 0\,--\,12\,Mm corresponds to north-to-south orientation in Figure\,\ref{fig:fov}.}
\label{fig:loopg}
\end{figure*}

In this appendix, the additional figures discussed in \S\,\ref{sec:res} are presented. In Figure\,\ref{fig:loopa}, the top panels (A1\,--\,A20) show the normalised \hic\ intensity slice (\textit{blue}) and the Gaussian profiles (\textit{grey}) that are used to generate the fit (\textit{red}). The cross-sectional profiles are each normalised to the maximum intensity of the loop segment. The bottom panel shows the cross-sectional width of the loop (\textit{black}) and the sub-element strands (\textit{cyan, magenta, green}) as a function of position along the loop segment. The error bars for the loop widths indicate $\pm1$\,\hic\ pixel whilst the error bars of the sub-element strand widths are the one\,-\,$\sigma$ errors returned from the curve-fitting method. The same plots are shown for loops C (Figure\,\ref{fig:loopc}), D (Figure\,\ref{fig:loopd}), E (Figure\,\ref{fig:loope}), F (Figure\,\ref{fig:loopf}), and G (Figure\,\ref{fig:loopg}). As can be seen from these figures, magnetically closed loops, D, F, and G, and open fan loops A and C show little variance in the cross-sectional loop width \textit{vs.} length, whilst open fan loop E undergoes significant expansion.

\end{appendix}


\end{document}